# ETHICS-BASED AUDITING OF AUTOMATED DECISION-MAKING SYSTEMS: NATURE, SCOPE, AND LIMITATIONS

Jakob Mökander[1], Jessica Morley[1], Mariarosaria Taddeo[1,2], Luciano Floridi[1,2]

[1] Oxford Internet Institute, University of Oxford, 1 St Giles', Oxford, OX1 3JS
[2] Alan Turing Institute, British Library, 96 Euston Rd, London NW1 2DB

<jakob.mokander@oii.ox.ac.uk>

**Abstract**
Important decisions that impact human lives, livelihoods, and the natural environment are increasingly being automated. Delegating tasks to so-called *automated decision-making systems* (ADMS) can improve efficiency and enable new solutions. However, these benefits are coupled with ethical challenges. For example, ADMS may produce discriminatory outcomes, violate individual privacy, and undermine human self-determination. New governance mechanisms are thus needed that help organisations design and deploy ADMS in ways that are ethical, while enabling society to reap the full economic and social benefits of automation. In this article, we consider the feasibility and efficacy of *ethics-based auditing* (EBA) as a governance mechanism that allows organisations to validate claims made about their ADMS. Building on previous work, we define EBA as a structured process whereby an entity's present or past behaviour is assessed for consistency with relevant principles or norms. We then offer three contributions to the existing literature. First, we provide a theoretical explanation of how EBA can contribute to good governance by promoting procedural regularity and transparency. Second, we propose seven criteria for how to design and implement EBA procedures successfully. Third, we identify and discuss the conceptual, technical, social, economic, organisational, and institutional constraints associated with EBA. We conclude that EBA should be considered an integral component of multifaced approaches to managing the ethical risks posed by ADMS.

# 1. INTRODUCTION

## 1.1. Background

Automated decision-making systems (ADMS), i.e. autonomous self-learning systems that gather and process data to make qualitative judgements with little or no human intervention, increasingly permeate all aspects of society (AlgorithmWatch 2019). This means that many decisions with significant implications for people and their environments – which were previously made by human experts – are now made by ADMS (Zarsky 2016; Karanasiou and Pinotsis 2017; Krafft et al. 2020). Examples of the use of ADMS by both governments and private entities include potentially sensitive areas like medical diagnostics (Grote and Berens 2020), recruitment (Sánchez-Monedero et al. 2020), driving autonomous vehicles (Evans et al. 2020), and the issuing of loans and credit cards (Aggarwal 2019). As information societies mature, the range of decisions that can be automated in this fashion will increase, and ADMS will be used to make ever-more critical decisions.

From a technical perspective, the specific models used by ADMS vary from simple decision trees to deep neural networks (Lepri et al. 2018). In this paper, however, we focus not on the underlying technologies but rather on the common features of ADMS from which ethical challenges arise. In particular, it is the combination of relative *autonomy*, *complexity*, and *scalability* that underpin both beneficial and problematic uses of ADMS (more on this in section 2). Delegating tasks to ADMS can help increase consistency, improve efficiency, and enable new solutions to complex problems (Taddeo and Floridi 2018). Yet these improvements are coupled with ethical challenges. As noted already by Norbert Wiener (1988 [1954]): "The machine, which can learn and can make decisions on the basis of its learning, will in no way be obliged to make such decisions as we should have made, or will be acceptable to us." For example, ADMS may leave decision subjects vulnerable to harms associated with poor-quality outcomes, bias and discrimination, and invasion of privacy (Leslie 2019). More generally, ADMS risk enabling human wrongdoing, reducing human control, removing human responsibility, devaluing human skills, and eroding human self-determination (Tsamados et al. 2020).

If these ethical challenges are not sufficiently addressed, a lack of public trust in ADMS may hamper the adoption of such systems which, in turn, would lead to significant social opportunity costs through the underuse of available and well-designed technologies (Cookson 2018). Addressing the ethical challenges posed by ADMS is therefore becoming a prerequisite for good governance in information societies (Cath et al. 2018). Unfortunately, traditional governance mechanisms designed to oversee human decision-making processes often fail when applied to ADMS (Kroll et al. 2016). One important reason for this is that the delegation of tasks to ADMS curtails the sphere of ethical deliberation in decision-making processes (D'Agostino and Durante 2018). In practice, this means that norms that used to be open for interpretation by human decision-makers are now embodied in ADMS. From an ethical perspective, this shifts the focus of ethical deliberation from specific decision-making situations to the ways in which ADMS are designed and deployed.

## 1.2. From principles to practice

In response to the growing need to design and deploy ADMS in ways that are ethical, over 75 organisations – including governments, companies, academic institutions, and NGOs – have produced documents defining high-level guidelines (Jobin et al. 2019). Reputable contributions include *Ethically Aligned Design* (IEEE 2019), *Ethics Guidelines for*

*Trustworthy AI* (AI HLEG 2019), and the OECD's *Recommendation of the Council on Artificial Intelligence* (OECD 2019). Although varying in terminology, the different guidelines broadly converge around five principles: beneficence, non-maleficence, autonomy, justice, and explicability (Floridi and Cowls 2019).

While a useful starting point, these principles tend to generate interpretations that are either too semantically strict, which are likely to make ADMS overly mechanical, or too flexible to provide practical guidance (Arvan 2018). This indeterminacy hinders the translation of ethics principles into practices and leaves room for unethical behaviours like 'ethics shopping', i.e. mixing and matching ethical principles from different sources to justify some pre-existing behaviour; 'ethics bluewashing', i.e. making unsubstantiated claims about ADMS to appear more ethical than one is; and 'ethics lobbying', i.e., exploiting ethics to delay or avoid good and necessary legislation (Floridi 2019). Moreover, the adoption of ethics guidelines remains voluntary, and the industry lacks both incentives and useful tools to translate principles into verifiable criteria (Raji et al. 2020). For example, interviews with software developers indicate that while they consider ethics important in principle, they also view it as an impractical construct that is distant from the issues they face in daily work (Vakkuri et al. 2019). Further, even organisations that are acutely aware of the risks posed by ADMS may struggle to manage these, either due to a lack of useful governance mechanisms or conflicting interests (PwC 2019). Taken together, there still exists a gap between the 'what' (and 'why') of ethics principles, and the 'how' of designing, deploying, and governing ADMS in practice (Morley et al. 2020).

A vast range of governance mechanisms that aim to support the translation of high-level ethics principles into practical guidance has been proposed in the existing literature. Some of these governance mechanisms focus on interventions in the early stages of software development processes, e.g. by raising the awareness of ethical issues among software developers (Floridi et al. 2018), creating more diverse teams of software developers (Sánchez-Monedero et al. 2020), embedding ethical values into technological artefacts through proactive design (Aizenberg and van den Hoven 2020; van de Poel 2020), screening potentially biased input data (AIEIG 2020), or verifying the underlying decision-making models and code (Dennis et al. 2016). Other proposed governance mechanisms, such as impact assessments (ECP 2018), take the outputs of ADMS into account. Yet others focus on the context in which ADMS operate. For example, so-called Human-in-the-Loop protocols imply that human operators can either intervene to prevent or be held responsible for harmful system outputs (Rahwan 2018; Jotterand and Bosco 2020).

**1.3. Scope, limitations, and outline**

One governance mechanism that merits further examination is ethics-based auditing (EBA) (Diakopoulos 2015; Raji et al. 2020; Brown et al. 2021). Operationally, EBA is characterised by a structured process whereby an entity's present or past behaviour is assessed for consistency with relevant principles or norms (Brundage et al. 2020). The main idea thereby is that the causal chain behind decisions made by ADMS can be revealed by improved procedural transparency and regularity, which, in turn, allow stakeholders to identify who should be held accountable for potential ethical harms. However, rather than attempting to codify ethics, EBA helps identify, visualise, and communicate whichever normative values are embedded in a system – with the aim of sparking ethical deliberation amongst software developers and managers in organisations that design and deploy ADMS. It is important to note that while EBA can provide useful and relevant information, they do not tell human decision-makers how to act on that information. That said, by strengthening trust between different stakeholders and promoting transparency, EBA can facilitate morally good actions (more on this in section 3).

The idea of auditing software is not new. Since the 1970s, computer scientists have been researching how to ensure that different software systems adhere to predefined functionality and reliability standards (Weiss 1980). Nor is the idea of auditing ADMS for consistency with ethics principles new. In 2014, Sandvig et al. referred to 'auditing of algorithms' as a promising, yet underexplored, governance mechanism to address the ethical challenges posed by ADMS. Since then, EBA has attracted much attention from policymakers, researchers, and industry practitioners alike. For example, regulators like the UK Information Commissioner's Office (ICO) have drafted AI auditing frameworks (ICO 2020; Kazim et al. 2021). At the same time, traditional accounting firms, including PwC (2019) and Deloitte (2020), technology-based startups like ORCAA (2020), and all-volunteer organisations like ForHumanity (2021) are all developing tools to help clients verify claims about their ADMS. However, despite a growing interest in EBA from both policymakers and private companies, important aspects of EBA are yet to be substantiated by academic research. In particular, a theoretical foundation for explaining how EBA affords good governance has hitherto been lacking.

In this article, we attempt to close this knowledge gap by analysing the *feasibility* and *efficacy* of EBA as a governance mechanism that allows organisations to operationalise their ethical commitments and validate claims made about their ADMS. Potentially, EBA can also serve the purpose of helping individuals understand how a specific decision was made as well as how to contest it. Our primary focus, however, is on the affordances and constraints of EBA as an organisational governance mechanism. The purpose thereby is to contribute to an improved understanding of what EBA is and how it can help organisations develop and deploy ethically-sound ADMS in practice.

To narrow down the scope of our analysis, we introduced two further limitations. First, we do not address any legal aspects of auditing. Rather, our focus in this article is on ethical alignment, i.e. on what ought and ought not to be done over and above compliance with existing regulation. This is not to say that hard governance mechanisms (like laws and regulations) are superfluous. In contrast, as stipulated by the AI HLEG (2019), ADMS should be lawful, ethical, and technically robust. However, hard and soft governance mechanisms often complement each other, and decisions made by ADMS can be ethically problematic and deserving of scrutiny even when not illegal (Floridi 2018). Hence, from now on, 'EBA' is to be understood as a soft yet formal[1] 'post-compliance' governance mechanism.

Second, any review of normative ethics frameworks remains outside the scope of this article. When designing and operating ADMS, tensions may arise between different ethical principles for which there are no fixed solutions (Kleinberg et al. 2017). For example, a particular ADMS may improve the overall accuracy of decisions but discriminate against specific subgroups in the population (Whittlestone et al. 2019a). Similarly, different definitions of fairness – like individual fairness, demographic parity, and equality of opportunity – are mutually exclusive (Kusner et al. 2017; Friedler et al. 2016). In short, it would be naïve to suppose that we have to – or indeed even can – resolve disagreements in moral and political philosophy (see e.g. Binns 2018) before we start to design and deploy ADMS. To overcome this challenge, we conceptualise EBA as a governance mechanism that can help organisations adhere to any predefined set of (coherent and justifiable) ethics principles (more on this in section 7.1). EBA can, for example, take place within one of the ethical frameworks already mentioned, especially the *Ethics Guidelines for Trustworthy AI* (AI HLEG 2019) for countries belonging to the European Union and the *Recommendation of the Council on Artificial Intelligence* (OECD 2019) for countries that officially adopted the OECD principles. But organisations that design and deploy ADMS may also formulate their own sets of ethics principles and use these as a baseline to audit. The main takeaway here is that EBA is not morally good in itself, nor it is sufficient to guarantee morally good outcomes. EBA enables moral goodness to be realised, if properly implemented and combined with justifiable values and sincere intentions (Floridi 2017a; Taddeo 2016).

---

[1] Formal (as opposed to informal) governance mechanisms are officially stated, documented, and communicated by the organisation that employs them (Falkenberg and Herremans 1995).

The remainder of this article proceeds as follows. In section 2, we define 'ADMS' and discuss the central features of ADMS that give rise to ethical challenges. In section 3, we explain what EBA is (or should be) in the context of ADMS. In doing so, we also clarify the roles and responsibilities of different stakeholders in relation to the EBA procedures. In section 4, we provide an overview of currently available frameworks and tools for EBA of ADMS and how are these being implemented. We then offer three main contributions to the existing literature. First, in section 5, we articulate how EBA can support good governance. Second, in section 6, we identify seven criteria for how to implement EBA procedures successfully. Third, in section 7, we highlight and discuss the constraints associated with EBA of ADMS. In section 8, we conclude that EBA, as outlined in this article, can help organisations manage some of the ethical risks posed by ADMS while allowing societies to reap the economic and social benefits of automation.

## 2. AUTOMATED DECISION-MAKING SYSTEMS

While 'algorithms', 'AI' and 'ADMS' are often used interchangeably, we prefer to use the term ADMS because it captures more precisely the essential features of the systems under investigation. Throughout this article, we are using the following definition of ADMS provided by AlgorithmWatch in their report *Automating Society (2019)*.

> "[Automatic Decision-Making System] refers to sociotechnical systems that encompass a decision-making model, an algorithm that translates this model into computable code, the data this code uses as an input, and the entire environment surrounding its use".

From an ethical perspective, it is primarily the *autonomous*, *complex*, and *scalable* nature of ADMS that either introduces new types of challenges or exacerbates existing societal tensions and inequalities. Although interlinked and mutually reinforcing, these three features pose different conceptual challenges. The autonomous nature of ADMS makes it difficult to predict and assign accountability when harms occur (Tutt 2017; Coeckelbergh 2020). Traditionally, the actions of technical systems have been linked to the user, the owner, or the manufacturer of the system. However, the ability of ADMS to adjust their behaviour over time undermines existing chains of accountability (Dignum 2017). Moreover, it is increasingly possible that a network of agents – some human, some non-human – may cause morally loaded actions (Floridi 2016a). The appearance of ADMS thereby challenges notions of moral agents as necessarily human in nature (Floridi 2013).

Similarly, the complex, often opaque, nature of ADMS may hinder the possibility of linking the outcome of an action to its causes (Oxborough et al. 2018). For example, the structures that enable learning in neural networks, including the use of hidden layers, contributes to technical opacity that may undermine the attribution of accountability for the action of ADMS (Citron and Pasquale 2014). While it should be noted that opacity can also be a result of intentional corporate or state secrecy (Burrell 2016), our main concern here relates to inherent technical complexity.

Finally, the scalability of ADMS implies that it will become more difficult to manage system externalities, as they may be hard to predict and spill across borders and generations (Dafoe 2017). This makes it challenging to define and reconcile different legitimate values and interests. The problem posed by the scalability of ADMS is thus not only that norms will become harder to uphold but also harder to agree upon in the first place. Of course, the levels of autonomy, complexity, and scalability displayed by ADMS are all matters of degree (Tasioulas 2018). For example, in some cases, ADMS act in full autonomy, whereas in others, ADMS provide recommendations to a human operator who has the final say (Cummings 2004). In terms of complexity, a similar distinction can be made between ADMS that automate routine tasks and those which learn from their environment to achieve goals.

From a governance perspective, it is useful to view highly autonomous and self-learning ADMS as parts of larger sociotechnical systems. Because ADMS adapt their behaviour based on external inputs and interactions with their environments (van de Poel 2020), important dynamics of the system as a whole may be lost or misunderstood if technical subsystems are targeted separately (Di Maio 2014). This risk is summarised by what Lauer (2020) calls the fallacy of the broken part: when there is a malfunction, the first instinct is to identify and fix the broken part. Yet most serious errors or accidents associated with ADMS can be traced not to coding errors but requirement flaws (Leveson 2011). This implies that no purely technical solution will be able to ensure that ADMS are ethically-sound (Kim 2017). It also implies that audits need to consider not only the source code of an ADMS and the purpose for which it is employed, but also the actual impact it exerts on its environment as well as the normative goals of relevant stakeholders.

## 3. ETHICS-BASED AUDITING

EBA is a *governance mechanism* that can be used by organisations to control or influence the ways in which ADMS are designed and deployed, and thereby, indirectly, shape the resultant characteristics of these systems (Mökander and Floridi 2021). As mentioned in the introduction, EBA is characterised by a *structured process* whereby an entity's present or past behaviour is assessed for consistency with relevant principles or norms (Brundage et al. 2020). It is worth noting that the entity in question, i.e. the subject of the audit, can be a person, an organisational unit, or a technical system. Taking a holistic approach, we argue that these different types of EBA are complementary.

Further, we use the expression 'ethics-based' instead of 'ethical' to avoid any confusion: We do neither refer to a kind of auditing conducted ethically, nor to the ethical use of ADMS in auditing, but to an auditing process that assesses ADMS based on their adherence to predefined ethics principles. Thus, EBA shifts the focus of the discussion from the abstract to the operational, and from guiding principles to managerial intervention throughout the product lifecycle, thereby permeating the conceptualisation, design, deployment and use of ADMS.

While widely accepted standards for EBA of ADMS have yet to emerge, it is possible to distinguish between different approaches. For example, *functionality audits* focus on the rationale behind decisions; *code audits* entail reviewing the source code of an algorithm; and *impact audits* investigate the types, severity, and prevalence of effects of an algorithm's outputs. Again, these approaches are complementary and can be combined to design and implement EBA procedures in ways that are feasible and effective (more on this in section 5.1).

Importantly, EBA differs from merely publishing a code of conduct, since its central activity consists of demonstrating adherence to a predefined baseline (ICO 2020). EBA also differs from certification in important aspects. For example, whereas certification typically aims at producing an official document that attests to a particular status or level of achievement (Scherer 2016). To this end, certifications are issued by a third party, whereas auditing can (in theory) be done by (parts of) an organisation over itself for purely internal purposes. In sum, understood as a process of informing, interlinking, and assessing existing governance structures, EBA can provide the basis for – but is not reducible to – certification.

As a governance mechanism that aims to promote trust and transparency, 'auditing' (as commonly understood) has a long history in areas like security and financial accounting (LaBrie and Steinke 2019). Valuable lessons can be learned from these domains. Most importantly, the process of 'auditing' is always purpose-oriented. In our case, EBA is directed towards the goal of ensuring that ADMS operate in ways that align with specific ethics guidelines. Throughout this purpose-oriented process, various *tools* (such as software programs and standardised reporting formats) and *methods* (like stakeholder consultation or adversarial testing) are employed to verify claims and create traceable documentation. This documentation process enables the identification of the reasons why an ADMS was erroneous, which, in turn, could help anticipate undesired consequences for certain stakeholders and prevent future mistakes (Felzmann et al. 2020). Naturally, different EBA procedures employ different tools and contain different steps. The protocols that govern specific EBA procedures are hereafter referred to as EBA *frameworks*.

Another lesson is that 'auditing' presupposes operational independence between the auditor and the auditee. Whether the auditor is a government body, a third-party contractor, an industry association, or a specially designated function within larger organisations, the main point is to ensure that the audit is run independently from the regular chain of command within organisations (Power 1999). The reason for this is not only to minimise the risk of collusion between auditors and auditees but also to clarify roles so as to be able to allocate responsibility for different types of harm or system failures (IIA 2017).

Figure 1 below illustrates the relationships between organisations that design and deploy ADMS (who are accountable for the behaviour of their systems), the management of such organisations (who are responsible for achieving organisational goals, including adhering to ethical values), the independent auditor (who is tasked with objectively reviewing and assessing how well an organisation adheres to relevant principles and norms), and regulators (who are monitoring the compliance of organisations on behalf of the government and decision-making subjects). For EBA to be effective, auditors must be able to test ADMS for a wide variety of typical and atypical scenarios. Regulators can therefore support the emergence and implementation of voluntary EBA procedures by providing the necessary infrastructure to share information and create standardised reporting formats and evaluation criteria (Keyes et al. 2019).

*Figure 1. Roles and responsibilities during independent audits*

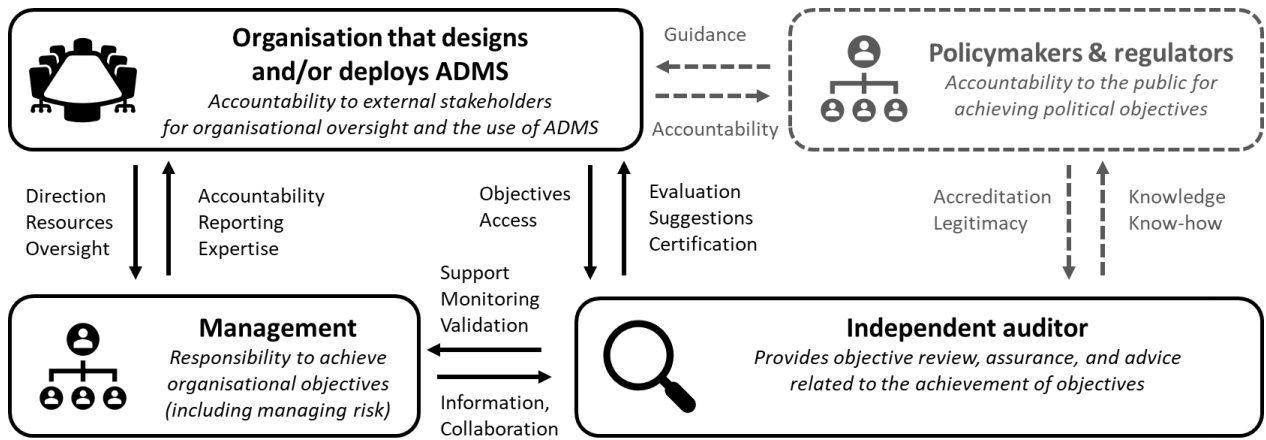

## 4. STATUS QUO: EXISTING EBA FRAMEWORKS AND TOOLS

In this section, we survey the landscape of currently available EBA frameworks and tools. In doing so, we illustrate how EBA can provide new ways of detecting, understanding, and mitigating the unwanted consequences of ADMS.

### 4.1. Ethics-based auditing frameworks

As described in the previous section, EBA *frameworks* are protocols that describe a specific EBA procedure and define what is to be audited, by whom, and according to which standards. Typically, EBA *frameworks* originate from one of two processes. The first type consists of 'top-down' national or regional strategies, like those published by the Government of Australia (Dawson et al. 2019) or Smart Dubai (2019). These strategies tend to focus on legal aspects or stipulate normative guidelines.[2]

At a European level, the debate was shaped by the AI4People project, which proposed that 'auditing mechanisms' should be developed to identify unwanted ethical consequences of ADMS (Floridi et al. 2018). Since then, the AI HLEG[3] has published not only the *Ethics-Guidelines for Trustworthy AI* (2019), but also a corresponding *Assessment List for Trustworthy AI* (2020). This assessment list is intended for self-evaluation purposes and can thus be incorporated into EBA procedures. Such checklists are simple tools that help designers get a more informed view of edge cases and system failures (Raji et al. 2020). Most recently, the European Commission (2021) published its long-anticipated proposal of the new EU *Artificial Intelligence Act*. The proposed regulation takes a risk-based approach. For our purposes, this means that a specific ADMS can be classified into one of four risk levels. While ADMS that pose 'unacceptable risk' are proposed to be completely banned, so-called 'high-risk' systems will be required to undergo legally mandated ex-ante and ex-post conformity assessments. However, even for ADMS that pose 'minimal' or 'limited' risk, the European Commission encourages organisations that design and deploy such systems to adhere to voluntary codes of conduct. In short, with respect to the proposed European regulation, there is a scope for EBA to help both providers of ADMS that pose limited risk to meet basic transparency obligations and providers of high-risk systems to demonstrate adherence to organisational values that goes over and above what is legally required.

The second type of EBA frameworks emerges 'bottom-up', from the expansion of data regulation authorities to account for the effects ADMS have on informational privacy. Building on an extensive experience of translating ethical principles into governance protocols, frameworks developed by data regulation agencies provide valuable blueprints for EBA of ADMS. The CNIL privacy impact assessment, for example, requires organisations to describe the context of the data processing under consideration when analysing how well procedures align with fundamental ethical principles (CNIL 2019). This need for contextualisation applies not only to data management but also to the use of ADMS at large. Another transferable lesson is that organisations should conduct an independent ethical evaluation of software they procure from – or outsource production to – third-party vendors (ICO 2018). At the same time, EBA frameworks with roots in data regulation tend to account only for specific ethical concerns, e.g. those related to privacy. This calls for caution. Since there is a plurality of ethical values which may serve as legitimate normative ends (think of freedom, equality, justice, proportionality, etc), an exclusive focus on one, or even a few, ethical challenges risks leading to sub-optimisation from a holistic perspective.

---

[2] For a review of existing sets of ethics principles and national or regional AI governance frameworks, interested readers are directed to either *The Ethics of AI Ethics* (Hagendorff 2020) or (Floridi and Cowls 2019).

[3] The AI HLEG is an independent expert group set up by the European Commission in June 2018.

To synthesise, the reviewed EBA frameworks converge around a procedure based on impact assessments. IAF (2019) summarised this procedure in eight steps: (1) Describe the purpose of the ADMS; (2) Define the standards or verifiable criteria based on which the ADMS should be assessed; (3) Disclose the process, including a full account of the data use and parties involved; (4) Assess the impact the ADMS has on individuals, communities, and its environment; (5) Evaluate whether the benefits and mitigated risks justify the use of ADMS; (6) Determine the extent to which the system is reliable, safe, and transparent; (7) Document the results and considerations; and (8) Reflect and evaluate periodically, i.e. create a feedback loop.

**4.2. Ethics-based auditing tools**

EBA *tools* are conceptual models or software products that help measure, evaluate, or visualise one or more properties of ADMS. With the aim to enable and facilitate EBA of ADMS, a great variety of such tools have already been developed by both academic researchers and privately employed data scientists. While these tools typically apply mathematical definitions of principles like fairness, accountability and transparency to measure and evaluate the ethical alignment of ADMS (Keyes et al. 2019), different tools help ensure the ethical alignment of ADMS in different ways. A full review of all the tools that organisations can employ during EBA procedures would be beyond the scope of this article. Nevertheless, in what follows, we provide some examples of different types of tools that help organisations design and develop ethically-sound ADMS.[4]

Some tools facilitate the audit process by visualising the outputs of ADMS. FAIRVIS, for example, is a visual analytics system that integrates a subgroup discovery technique, thereby informing normative discussions about group fairness (Cabrera et al. 2019). Another example is Fairlearn, an open-source toolkit that treats any ADMS as a black box. Fairlearn's interactive visualisation dashboard helps users compare the performance of different models (Microsoft 2020). These tools are based on the idea that visualisation helps developers and auditors to create more equitable algorithmic systems.

Other tools improve the interpretability of complex ADMS by generating more straightforward rules that explain their predictions. For example, Shapley Additive exPlanations, or SHAP, calculates the marginal contribution of relevant features underlying a model's prediction (Leslie 2019). The explanations provided by such tools are useful, e.g. when determining whether protected features have unjustifiably contributed to a decision made by ADMS. However, such explanations also have important limitations. For example, tools that explain the contribution of features that have been intentionally used as decision inputs may not determine whether protected features have contributed unjustifiably to a decision through proxy variables.

Yet other tools help convey the reasoning behind ADMS by applying one of three strategies: *Data-based explanations* provide evidence of a model by using comparisons with other examples to justify decisions; *Model-based explanations* focus on the algorithmic basis of the system itself; and *Purpose-based explanations* focus on comparing the stated purpose of a system with the measured outcomes (Kroll 2018). For our purposes, the key takeaway is that, while different types of explanations are possible, EBA should focus on local interpretability, i.e. explanations targeted at individual stakeholders – such as decision subjects or external auditors – and for specific purposes like internal governance, reputation management, or third-party verification. Here, a parallel can be made to what Loi et al. (2020) call *transparency as design publicity*, whereby organisations that design or deploy ADMS are expected to publicise the intentional explanation of the use of a specific system as well as the procedural justification of the decision it takes.

---

[4] The EBA frameworks and tools reviewed in this section are summarised in Appendix A.

Tools have also been developed that help to democratise the study of ADMS. Consider the TuringBox, which was developed as part of a time-limited research project at MIT. This platform allowed software developers to upload the source code of an ADMS so as to let others examine them (Epstein et al. 2018). The TuringBox thereby provided an opportunity for developers to benchmark their system's performance with regards to different properties. Simultaneously, the platform also allowed independent researchers to evaluate the outputs from ADMS, thereby adding an extra layer of procedural transparency to the software development process.

Finally, some tools help organisations document the software development process and monitor ADMS throughout their lifecycle. AI Fairness 360 developed by IBM, for example, includes metrics and algorithms to monitor, detect, and mitigate bias in datasets and models (Bellamy et al. 2019). Other tools have been developed to aid developers in making pro-ethical design choices (Floridi 2016b) by providing useful information about the properties and limitations of ADMS. Such tools include end-user license agreements, tools for detecting bias in datasets (Saleiro et al. 2018), and tools for improving transparency like datasheets for datasets (Gebru et al. 2018).

## 5. A VISION FOR ETHICS-BASED AUDITING OF ADMS

### 5.1. Connecting the dots

As demonstrated in section 4 above, a wide variety of EBA frameworks and tools have already been developed to help organisations and societies manage the ethical risks posed by ADMS. However, these tools are often employed in isolation. Hence, to be feasible and effective, EBA procedures need to combine existing conceptual frameworks and software tools into a structured process that monitors each stage of the software development lifecycle to identify and correct the points at which ethical failures (may) occur. In practice, this means that EBA procedures should combine elements of (a) *functionality auditing,* which focuses on the rationale behind decisions (and why they are made in the first place); (b) *code auditing,* which entails reviewing the source code of an algorithm; and (c) *impact auditing,* whereby the severity and prevalence of the effects of an algorithm's outputs are investigated (Mittelstadt 2016).

It should be reemphasised that the primary responsibility for identifying and executing steps to ensure that ADMS are ethically sound rests with the management of the organisations that design and operate such systems. In contrast, the independent auditor's responsibility is to (i) assess and verify claims made by the auditee about its processes and ADMS and (ii) ensure that there is sufficient documentation to respond to potential inquiries from public authorities or individual decision subjects. More proactively, the process of EBA should also help spark and inform ethical deliberation throughout the software development process. The idea is that continuous monitoring and assessment ensures that a constant flow of feedback concerning the ethical behaviour of ADMS is worked into the next iteration of their design and application. Figure 2 below illustrates how the process of EBA runs in parallel with the software development lifecycle.

*Figure 2. EBA helps inform, formalise, and interlink existing governance structures through an iterative process*

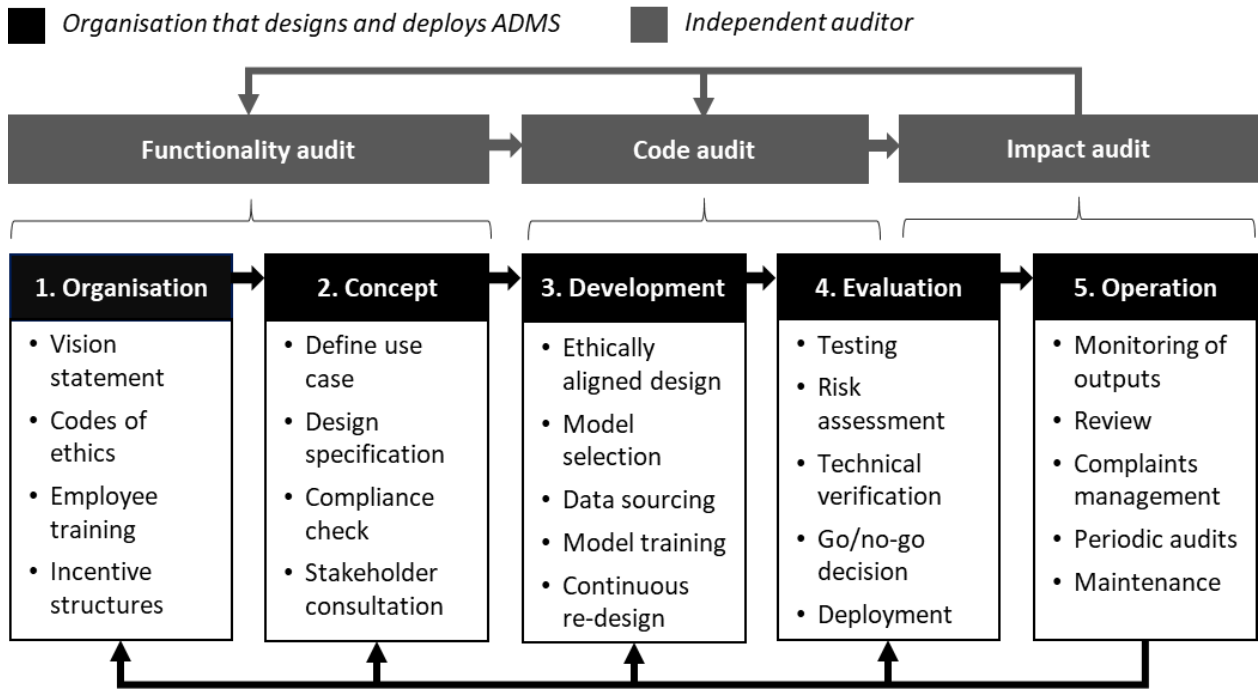

**5.2. Methodological advantages**

EBA of ADMS – as outlined in this article – displays six, interrelated and mutually reinforcing, methodological advantages. These are best illustrated by examples from existing tools:

1) EBA can provide decision-making support to executives and legislators by defining and monitoring outcomes, e.g. by showing the normative values embedded in a system (AIEIG 2020). Here, EBA serves a diagnostic function: before asking whether we would expect an ADMS to be ethical, we must consider which mechanisms we have to determine what it is doing at all. By gathering data on system states (both organisational and technical) and reporting on the same, EBA enables stakeholders to evaluate the reliability of ADMS in more detail. A systematic audit is thereby the first step to make informed model selection decisions and to understand the causes of adverse effects (Saleiro et al. 2018).

2) EBA can increase public trust in technology and improve user satisfaction by enhancing operational consistency and procedural transparency. Mechanisms such as documentation and actionable explanations are essential to help individuals understand why a decision was reached and contest undesired outcomes (Wachter et al. 2018). This also has economic implications. While there may be many justifiable reasons to abstain from using available technologies in certain contexts, fear and ignorance may lead societies to underuse available technologies even in cases where they would do more good than harm (Floridi et al. 2018). In such cases, increased public trust in ADMS could help unlock economic growth. However, to drive trust in ADMS, explanations need to be actionable and selective (Barredo Arrieta et al. 2020). This is possible even when algorithms are technically opaque since ADMS can be understood intentionally and in terms of their inputs and outputs.

3) EBA allows for local alignment of ethics and legislation. While some normative metrics must be assumed when evaluating ADMS, EBA is a governance mechanism that allows organisations to choose which set of ethics principles they seek to adhere to. This allows for contextualisation. Returning to our example with fairness above, the most important aspect from an EBA perspective is not which specific definition of fairness is applied in a particular case, but that this decision is communicated transparently and publicly justified. In short, by focusing on identifying errors, tensions, and risks, as well as communicating the same to relevant stakeholders, such as customers or independent industry associations, EBA can help organisations demonstrate adherence to both sector-specific and geographically dependent norms and legislation.

4) EBA can help relieve human suffering by anticipating potential negative consequences before they occur (Raji et al. 2020). There are three overarching strategies to mitigate harm: pre-processing, i.e. reweighing or modifying input data; in-processing, i.e. model selection or output constraints; and post-processing, i.e. calibrated odds or adjustment of classifications (Koshiyama 2019). These strategies are not mutually exclusive. By combining minimum requirements on system performance with automated controls, EBA can help both developers test and improve the performance of ADMS (Mahajan et al. 2020) and enable organisations to establish safeguards against unexpected or unwanted behaviours.

5) EBA can help balance conflicts of interest. A right to explanation must, for example, be reconciled with jurisprudence and counterbalanced with intellectual property law as well as freedom of expression (Wachter et al. 2018). By containing access to sensitive parts of the review process to authorised third-party auditors, EBA can provide a basis for accountability while preserving privacy and intellectual property rights.

6) EBA can help human decision-makers to allocate accountability by tapping into existing internal and external governance structures (Bartosch et al. 2018). Within organisations, EBA can forge links between non-technical executives and developers. Externally, EBA help organisations validate the functionality of ADMS. In short, EBA can clarify the roles and responsibilities of different stakeholders and, by leveraging the capacity of institutions like national civil courts, help to redress the harms inflicted by ADMS.

Naturally, the methodological advantages highlighted in this section are potential and far from being guaranteed. However, the extent to which these benefits can be harnessed in practice depends not only on complex contextual factors but also on how EBA frameworks are designed. To realise its full potential as a governance mechanism, EBA of ADMS needs to meet specific criteria. In the next section, we turn to specifying these criteria.

## 6. CRITERIA FOR SUCCESSFUL IMPLEMENTATION

Best practices for EBA of ADMS have yet to emerge. Nevertheless, as discussed in section 4, organisations and researchers have already developed, and attempted to pilot, a wide range of EBA tools and frameworks. These early attempts hold valuable and generalisable lessons for organisations that wish to implement feasible and effective EBA procedures. As we will see, some of these lessons concern how stakeholders view EBA of ADMS, whilst other lessons concern the design of EBA practices. In this section, we will discuss the most important lessons from previous work and condense these into criteria for how to get EBA of ADMS right.

As a starting point, it should be acknowledged that ADMS are not isolated technologies. Rather, ADMS are both shaped by and help shape larger sociotechnical systems (Dignum 2017). Hence, system output cannot be considered biased or erroneous without some knowledge of the available alternatives. Therefore, a holistic approach to EBA of ADMS needs to seek input from diverse stakeholders, e.g. for an inclusive discourse about key performance indicators (KPI). However, regardless of which KPI an organisation chooses to adopt, audits are only meaningful insofar as they allow organisations to verify claims made about their ADMS. This implies that EBA procedures themselves must be *traceable*. By providing a traceable log of the steps taken in the design and development of ADMS, audit trails can help organisations verify claims about their engineered systems (Brundage et al. 2020). However, a distinction should be made between traceability and transparency: while transparency is often invoked to improve trust (Springer and Whittaker 2019), full transparency concerning the content of audits may not be desirable, e.g. with regards to privacy- and intellectual property rights. Instead, what counts is procedural transparency and regularity.

Further, to ensure that ADMS are ethically-sound, organisational policies need to be broken down into tasks for which individual agents can be held *accountable* (Ananny and Crawford 2018). By formalising the software development process and revealing (parts of) the causal chain behind decisions made by ADMS, EBA helps clarify the roles and responsibilities of different stakeholders, including executives, process owners, and data scientists. However, allocating responsibilities is not enough. Sustaining a culture of trust also requires that people who breach ethical and social norms are subject to proportional sanctions (Ellemers et al. 2019). By providing avenues for whistle-blowers and promoting a culture of ethical behaviour, EBA also helps strengthen interpersonal accountability within organisations (Koene et al. 2019). At the same time, doing the right thing should be made easy. This can be achieved through strategic governance structures that align profit with purpose. The 'trustworthiness' of a specific ADMS is never just a question about technology but also about value alignment (Gabriel 2020; Christian 2020). In practice, this means that the checks and balances developed to ensure safe and benevolent ADMS must be incorporated into policies, organisational *strategies*, and individual paths.

Importantly, EBA does not provide an answer sheet but a playbook. This means that EBA of ADMS should be viewed as a *dialectic* process wherein the auditor ensures that the right questions are asked (Goodman 2016) and answered adequately. Auditors and systems owners work together to develop context-specific methods (Schulam and Saria 2019). To manage the risk that independent auditors would be too easy on their clients, licences should be revoked from both auditors and system owners in cases where ADMS fail. However, it is difficult to ensure that an ADMS contains no bias or to guarantee its fairness (Microsoft 2020). Instead, the goal from an EBA perspective should be to provide useful information about when an ADMS is causing harm or when it is behaving in a way that is different from what is expected. This pragmatic insight implies that audits need to monitor and evaluate system outputs *continuously*, i.e. through 'oversight programs' (Etzioni and Etzioni 2016), and document performance characteristics in a comprehensible way (Mitchell et al. 2019). Hence, continuous EBA of ADMS implies considering system impacts as well as organisations, people, processes, and products.

Finally, the alignment between ADMS and specific ethical values is a design question. Ideally, properties like interpretability and robustness should be built into systems from the start, e.g. through 'Value-Aligned Design' (Bryson and Winfield 2017). However, the context-dependent behaviour of ADMS makes it difficult to anticipate the impact ADMS will have on the complex environments in which they operate (Chopra and Singh 2018). By incorporating an active feedback element into the software development process, EBA can help inform the continuous re-design of ADMS. Although this may seem radical, it is already happening: most sciences, including engineering and jurisprudence, do not only study their systems, they simultaneously build and modify them (Floridi 2017b).

Taken together, these generalisable lessons suggest that even imperfectly implemented EBA procedures can make a real difference to the ways in which ADMS are designed and deployed. However, our analysis of previous work also finds that, in order to be feasible and effective, EBA procedures must meet *seven criteria*. More specifically, to help organisations manage the ethical risks posed by ADMS, we argue that EBA procedures should be:

1) *Holistic*, i.e. treat ADMS as an integrated component of larger sociotechnical contexts
2) *Traceable*, i.e. assign responsibilities and document decisions to enable follow-up
3) *Accountable*, i.e. help link unethical behaviours to proportional sanctions
4) *Strategic*, i.e. align ethical values with policies, organisational strategies, and incentives
5) *Dialectic*, i.e. view EBA as a constructive and collaborative process
6) *Continuous*, i.e. identify, monitor, evaluate, and communicate system impacts over time
7) *Driving re-design*, i.e. provide feedback and inform the continuous re-design of ADMS

Of course, these criteria are aspirational and, in practice, unlikely to be satisfied all at once. Nevertheless, we must not let perfect be the enemy of good. Policymakers and organisations that design and deploy ADMS are thus advised to consider these seven criteria when developing and implementing EBA procedures.

## 7. DISCUSSION: CONSTRAINTS ASSOCIATED WITH ETHICS-BASED AUDITING

Despite the methodological advantages identified in section 5, it is important to remain realistic about what EBA can, and cannot, be expected to achieve. Our analysis of existing tools and frameworks suggests that EBA of ADMS – even if implemented according to the criteria listed in section 6 – is subject to a range of conceptual, technical, social, economic, organisational, and institutional constraints. For an overview, please find these constraints summarised in table format in Appendix C. In the remainder of this section, we highlight and discuss the most pressing constraints associated with EBA of ADMS. To design feasible and effective EBA procedures, these constraints must be understood and accounted for.

### 7.1. Conceptual constraints

Conceptual constraints cannot be easily overcome by means of technical innovation or political decision. Instead, they must be managed continuously by balancing the need for ethical alignment with tolerance and respect for pluralism. Insofar as ethical guidelines often mask unresolved disputes about the definitions of normative concepts like fairness and justice, EBA of ADMS may be conceptually constrained by hidden political tensions. For example, the reviewed literature accommodates more than six definitions of fairness, including individual fairness, demographic parity, and equality of opportunity (Kusner et al. 2017). Some of these interpretations are mutually exclusive, and specific definitions of fairness can even increase discrimination according to others.

While EBA of ADMS can help ensure compliance with a given policy, how to prioritise between conflicting interpretations of ethical principles remains a normative question. This is because translating principles into practice often requires trade-offs between different legitimate, yet conflicting normative values. Using personal data, for example, may improve public services by tailoring them but compromise privacy. Similarly, while increased automation could make lives more convenient, it also risks undermining human autonomy. How to negotiate justifiable trade-offs is a context-dependent, multi-variable problem. While audits cannot guarantee that a justifiable balance has been struck, the identification, evaluation, and communication of trade-offs can be included as assessment criteria. One function of EBA is thus to make visible implicit choices and tensions, give voice to different stakeholders, and arrive at resolutions that, even when imperfect, are at least publicly defensible (Whittlestone et al. 2019b).

Moreover, EBA is constrained by the difficulty of quantifying externalities that occur due to indirect causal chains over time. This problem is exacerbated by the fact that the quantification of social phenomena inevitably strips away local knowledge and context (Mau and Howe 2019). On the one hand, tools claiming to operationalise ethics mathematically fall into the trap of technological solutionism (Lipton and Steinhardt 2019). On the other hand, tools that focus on only minimum requirements provide little incentives for organisations to go beyond the minimum.

### 7.2. Technical constraints

Technical constraints are tied to the autonomous, complex, and scalable nature of ADMS. These constraints are time and context-dependent and thus likely to be relieved or transformed by future research. Three of them are worth highlighting. First, consider how the complexity and the lack of transparency of machine learning models hinder their interpretation (Oxborough et al. 2018). Such characteristics of ADMS limit the effectiveness of audits insofar as they make it difficult to assign and trace responsibility when harm occurs. Technical complexity also makes it difficult to audit a system without perturbing it. Further, there is a risk that sensitive data may be exposed during the audit process itself (Kolhar et al. 2017). To manage this challenge, third party auditors can be given privileged and secured access to private information to assess whether claims about the safety, privacy, and accuracy made by the system developer are valid. As of today, however, most EBA schemes do not protect user data from third-party auditors.

A second technical constraint stems from the use of agile software development methods. The same agile qualities that help developers meet rapidly changing customer requirements also make it difficult for them to ensure compliance with pre-specified requirements. One approach to managing this tension is to incorporate agile methodologies (see e.g. Strenge and Schack 2020) that make use of 'living traceability' in the audit process. These methods provide snapshots of programs under development in real-time (Steghöfer et al. 2019). Despite the availability of such pragmatic fixes, however, the effectiveness of EBA remains limited by an inability to ensure the compliance of systems that are yet to emerge.

Finally, EBA is technically constrained by the fact that laboratories differ from real-life environments (Auer and Felderer 2018). Put differently, given the data- and context-dependent behaviour of ADMS, only limited reasoning about their later performance is possible based on testing in controlled settings. To manage this challenge, test environments for simulation can be complemented by continuous EBA of live applications which constantly execute the algorithm. One example is 'live experimentation', i.e. the controlled deployment of experimental features in live systems to collect runtime data and analyse the corresponding effect (Fagerholm et al. 2014). Still, meaningful quality assurance is not always possible within test environments.

## 7.3. Economic and social constraints

Economic and social constraints are those derived from the incentives of different actors. Unless these incentives are aligned with the normative vision for ethically-sound ADMS, economic and social constraints will reduce both the feasibility and effectiveness of EBA. Inevitably, EBA imposes costs, financial and otherwise. Even when the costs of audits are justifiable compared to the aggregated benefits, society will face questions about which stakeholders would reap which benefits and pay which costs. For example, the cost of EBA risks having a disproportionate impact on smaller companies (Goodman 2016). Similarly, licensing systems for ADMS are likely to be selectively imposed on specific sectors, like healthcare or air traffic (Council of Europe 2018). The point is that both the costs and benefits associated with EBA should be distributed to not unduly burden or benefit particular groups in, or sectors of, society. Similarly, demands for ethical alignment must be balanced with incentives for innovation and adoption. Pursuing rapid technological progress leaves little time to ensure that developments are robust and ethical (Whittlestone et al. 2019b). Thus, companies find themselves wedged between the benefits of disruptive innovation and social responsibility (Turner Lee 2018) and may not act ethically in the absence of oversight.

Moreover, there is always a risk of adversarial behaviour during audits. The ADMS being audited may, for example, attempt to trick the auditor (Rahwan 2018). An example of such behaviour was the diesel emission scandal, during which Volkswagen intentionally bypassed regulations by installing software that manipulated exhaust gases during tests (Conrad 2018). An associated risk is that emerging EBA frameworks end up reflecting and reinforcing existing power relations. Given an asymmetry in both know-how and computational resources between data controllers and public authorities, auditors may struggle to review ADMS (Kroll 2018). For example, industry representatives may choose not to reveal insider knowledge but instead use their informational advantage to obtain weaker standards (Koene et al. 2019). Sector-specific approaches may therefore lead to a shift of power and responsibility from juridical courts to private actors. Even if, in such a scenario, audits reveal flaws within ADMS, asymmetries of power may prevent corrective steps from being taken.

Another concern relates to the fact that ADMS increasingly mediate human interactions. From an EBA perspective, nudging, i.e. the process of influencing personal preferences through positive reinforcement or indirect suggestion (Thaler and Sunstein 2008), may shift the normative baseline against which ethical alignment is benchmarked. This risk is aggravated by 'automation bias', i.e. the tendency amongst humans to trust information that originates from machines more than their own judgement (Cummings 2004). Consequently, the potentially transformative effects associated with ADMS pose challenges for how to trigger and evaluate audits.

## 7.4. Organisational and institutional constraints

Organisational and institutional constraints concern the operational design of EBA frameworks. Because these constraints depend on legal sanctioning, they are inevitably linked to questions about power. Who audits whom? As of today, a clear institutional structure is lacking. To establish integrity and validity, EBA of ADMS must therefore adhere to a transparent and well-recognised process. However, both internal audits and those performed by professional service providers are subject to concerns about objectivity. A more plausible way to mandate EBA of ADMS would be the creation of a regulatory body to oversee system owners and auditors. Just as the Food and Drug Administration tests and approves medicines, a similar agency could be set up to approve specific types of ADMS (Tutt 2017). Such an agency would be able to engage in *ex ante* regulation rather than relying on *ex post* judicial enforcement. However, the main takeaway is that EBA will only be as good as the institution backing it (Boddington et al. 2017).

In a similar vein, EBA is only effective if auditors have access to the information and resources required to carry out rigorous and meaningful audits. Thus, EBA is infeasible without strong regulatory compulsion or cooperation from system owners. Data controllers have, for example, an interest not to disclose trade secrets. Moreover, the resources required to audit ADMS can easily exceed those available to auditors. If, for example, auditors have no information about special category membership, they cannot determine whether a disparate impact exists. Consequently, the effectiveness of EBA is constrained by a lack of access to both relevant information and resources in terms of manpower and computing power.

There are also fundamental tensions between national jurisdictions and the global nature of technologies (Erdelyi and Goldsmith 2018). Thus, rules need to be harmonised across domains and boarders. However, such efforts face a hard dilemma. On the one hand, the lack of shared ethical standards for ADMS may lead to protectionism and nationalism. On the other hand, policy discrepancies may cause a race to the bottom where organisations seek to establish themselves in territories that provide a minimal tax burden and maximum freedom for technological experimentation (Floridi 2019). As a result, the effectiveness of EBA of ADMS remains constrained by the lack of international coordination.

## 8. CONCLUSIONS

The responsibility to ensure that ADMS are ethically-sound lies with the organisations that develop and operate them. EBA – as outlined in this article – is a governance mechanism that helps organisations not only to ensure but also demonstrate that their ADMS adhere to specific ethics principles. Of course, this does not mean that traditional governance mechanisms are redundant. On the contrary, by contributing to procedural regularity and transparency, EBA of ADMS is meant to complement, enhance, and interlink other governance mechanisms like human oversight, certification, and regulation. For example, by demanding that ethics principles and codes of conduct are clearly stated and publicly communicated, EBA ensures that organisational practices are subject to additional scrutiny which, in turn, may counteract 'ethics shopping'. Similarly, EBA helps reduce the risk for 'ethics bluewashing' by allowing organisations to validate the claims made about their ethical conduct and the ADMS they operate. Thereby, EBA constitutes an integral component of multifaceted approaches to managing the ethical risks posed by ADMS.

In particular, continuous EBA can help address some of the ethical challenges posed by autonomous, complex, and scalable ADMS. However, even in contexts where EBA is necessary to ensure ethical alignment of ADMS, it is by no means sufficient. For example, it remains unfeasible to anticipate all long-term and indirect consequences of a particular decision made by an ADMS. Further, while EBA can help ensure alignment with a given policy, how to prioritise between irreconcilable normative values remains a fundamentally normative question. Thus, even if private initiatives to develop EBA mechanisms should be encouraged, one should resist the shift of power and ultimate responsibility from juridical courts to private actors. The solution here is that regulators should retain supreme sanctioning power by authorising independent agencies which, in turn, conduct EBA.

The constraints highlighted in this article do not seek to diminish the merits of EBA of ADMS. In contrast, our aim has been to provide a roadmap for future work. While all constraints listed constitute important fields of research, social concerns related to the potentially transformative effects of ADMS deserve specific attention. By shifting the normative base on which liberal democracy is built, ADMS may undermine this trust. Therefore, the design and implementation of EBA frameworks must be viewed as a part of – and not separated from – the debate about the type of society humanity wants to live in, and what moral compromises individuals are willing to strike in its making.

In conclusion, standardised EBA procedures can help organisations validate claims about their ADMS and help strengthen the institutional trust that is foundational for good governance. However, EBA will not and should not replace the need for continuous ethical reflection and deliberation among individual moral agents.

Mökander et al. (2021) Pre-print. Accepted for publication in the journal *Science and Engineering Ethics* (Springer)**Appendix A – List of reviewed EBA frameworks and tools**

Table 1 below summarised the EBA tools and frameworks reviewed in section 4.1. The Table thereby (non-exhaustively) lists some of the most recent and important contributions to developing EBA of ADMS (for the methodology used to produce this Table, see Appendix B).

*Table 1. List of reviewed EBA frameworks (F) and tools (T)*

| *Institution* | *Publication* | *Type* | *Source* |
| --- | --- | --- | --- |
| AI Ethics Impact Group | Framework to operationalise AI | F | (AIEIG 2020) |
| CNIL (France) | Privacy Impact Assessment | F | (CNIL 2019) |
| ECP (Netherlands) | AI Impact Assessment | F | (ECP 2018) |
| European Commission | Guidelines for trustworthy AI | F | (AI HLEG 2019) |
| Gov. of Australia | AI: Australia's Ethics Framework | F | (Dawson et al. 2019) |
| Gov. of Canada | Algorithmic Impact Assessment | F | (Gov. of Canada 2019) |
| ICO (UK) | AI auditing framework (Guidance) | F | (ICO 2020) |
| PDPC (Singapore) | Model AI Governance Framework | F | (PDPC 2020) |
| Smart Dubai (UAE) | AI Ethics Principles & Guidelines | F | (Smart Dubai 2019) |
| WEF | Facial Recognition Assessment | F | (WEF 2020) |
| CMU | FAIRVIS | T | (Cabrera et al. 2019) |
| Google | What-if-tool | T | (Google 2020) |
| IBM | AI Fairness 360 | T | (Bellamy et al. 2019) |
| Microsoft | Fairlearn | T | (Microsoft 2020) |
| MIT | TuringBox | T | (Epstein et al. 2018) |
| PwC | Responsible AI Toolkit | T | (PwC 2019) |
| University of Chicago | Aequitas | T | (Saleiro et al. 2018) |
| University of Texas | CERTIFAI | T | (Sharma et al 2019) |

**Appendix B – Methodology**

As mentioned in the introduction, the purpose of this article was to contribute to an improved understanding of what EBA is and how it can help organisations develop and deploy ethically-sound ADMS in practice. To achieve this aim, we let the following three questions guide the research that led up to this article:

1) What EBA tools and frameworks are currently available to ensure ethical alignment of ADMS, and how are these being implemented?
2) How can EBA of ADMS help organisations and society reap the full benefit of new technologies while mitigating the ethical risks associated with ADMS?
3) What are the conceptual, technical, economic, social, organisational, and institutional constraints associated with auditing of ADMS?

Questions (1)-(3) are listed in logical order, but chronologically (2) takes priority, so we began with a systematised review of existing literature to address (2). The collection phase involved searching five databases (Google Scholar, Scopus, SSRN, Web of Science and arXiv) for articles related to auditing of ADMS. Keywords for the search included ('auditing, 'evaluation', OR 'assessment') AND ('ethics', 'fairness', transparency', OR 'robust) AND ('automated decision-making', 'artificial intelligence', OR 'algorithms'). To limit the scope of the literature review, we focused on articles published after 2011, the year when IBM Watson marked the coming of the second wave of AI by beating the two best-ever humans to have competed in the TV quiz show Jeopardy (Susskind and Susskind 2015). In total, 122 articles and reports were included in the systematised literature review.

In a second step, existing auditing tools and frameworks were reviewed and evaluated to extract generalisable lessons to address (1) and then (3) by identifying the opportunities and constraints associated with implementing auditing of ADMS in practice. The tools and frameworks reviewed for this article, see Table 1 in appendix A, were selected on the virtue of being recent, relevant, and developed by reputable organisations.

**Appendix C – Summary table of constraints associated with EBA of ADMS**

As emphasises in section 7, EBA of ADMS is subject to a range of conceptual, technical, social, economic, organisational, and institutional constraints. These are summarised in Table 2[5] below. To design feasible and effective auditing procedures, these constraints must be understood and accounted for. Our hope is therefore that the constraints listed below will provide a roadmap for future research, and guide policymakers attempts to support emerging EBA practices.

*Table 2. Summary of constraints associated with EBA of ADMS*

| *Type* | *Constraints* |
|---|---|
| Conceptual | Lack of consensus around high-level ethical principles |
| | Normative values conflict and require trade-offs |
| | It is difficult to quantify externalities of complex systems |
| | Information is infallibly lost through reductionist explanations |
| Technical | Complex systems appear opaque and are hard to interpret |
| | Data integrity and privacy are exposed to risks during audits |
| | Linear compliance mechanisms are incompatible with agile development |
| | Tests may not be indicative of ADMS behaviour in real-world environments |
| Economic and Social | Auditing may disproportionately disadvantage specific sectors or groups |
| | Ensuring ethical alignment must be balanced with incentives for innovation |
| | Audits are vulnerable to adversarial behaviour |
| | The transformative effects of ADMS challenge notions of human dignity |
| | Emerging audit frameworks reflect and reinforce existing power relations |
| Organisational and Institutional | There is a lack of institutional clarity about who audits whom |
| | Auditors may lack the access or information required to evaluate ADMS |
| | The global nature of ADMS challenge national jurisdictions |

---

[5] This summary table was first published in our short commentary article (Mökander and Floridi 2021).